\documentclass[aps,prd,twocolumn,superscriptaddress,floatfix]{revtex4-1}

\usepackage{graphicx, subfigure}
\usepackage{amsmath}
\usepackage{amsfonts}
\usepackage{amssymb}
\usepackage{slashed}
\usepackage{hyperref}
\usepackage{amsbsy}
\usepackage{natbib}
\usepackage{balance}
\usepackage[utf8]{inputenc}
\usepackage{cleveref}
\usepackage{enumitem}

\newcommand{\be}{\begin{equation}}
\newcommand{\ee}{\end{equation}}
\newcommand{\ba}{\begin{eqnarray}}
\newcommand{\ea}{\end{eqnarray}}

\setlist[itemize]{leftmargin=0.5cm}
\begin{document}

\title{On the gradient flow of the Lagrangian density in non-Abelian gauge theories}

\author{E. Pallante}
\affiliation{Van Swinderen Institute for Particle Physics and Gravity, University of Groningen, 9747 AG, The Netherlands}


\begin{abstract}
We observe that the would-be running coupling on the lattice defined by means of the gradient-flow method in order to identify the conformal window of QCD is not renormalization-group invariant (RGI).
Indeed, we show that  the would-be running coupling, $g_{wb}^2(t)\!\propto\! t^2\langle E(t)\rangle$,    
-- with $\langle E(t)\rangle$ the expectation value of the Lagrangian density, ${\mbox{Tr}}G^2$, smeared on a radius $\sqrt{t}$ by means of the gradient flow --  has an anomalous dimension associated to the multiplicative renormalization factor of $t^2\langle E(t)\rangle$.
As a consequence, at a nontrivial infrared (IR) fixed point with nonvanishing anomalous dimension, $\gamma_*$, in the conformal window, the would-be running coupling vanishes asymptotically as $g_{wb}^2(t)\!\propto\!t^2\langle E(t)\rangle\!\sim\! t^{-\gamma_*/2}$ and does not scale as $g_{wb}^2(t)\!\propto\! t^2\langle E(t)\rangle\!\sim\!g_{wb}^{*2}\!\neq\!0$, with $g_{wb}^{*}$ the nonvanishing would-be coupling at the nontrivial fixed point, as postulated in the literature. The associated would-be beta function, 
$\beta_{wb}({g}_{wb}^2(t))$, is not proportional to a true RGI beta function, and it also vanishes asymptotically in the IR as $\beta_{wb}\!\sim\!\gamma_*g_{wb}^2(t)$ for nonvanishing $\gamma_*$ at the IR fixed point. Moreover, $\beta_{wb}$ violates two-loop universality and may develop spurious zeroes  both in the confined phase and the conformal window, despite $g_{wb}^2(t)$ is asymptotic to a true RGI running coupling in a neighborhood of the asymptotically free fixed point.
Our analysis allows us to reinterpret the  contradictory lattice results based on this method, specifically those for the $N_f\!=\!12$ theory, explain the origin of their discrepancies and suggest a new strategy to discriminate between the confined phase and the conformal window. In this respect, we disagree with a recent claim that attributes the same contradictory results to staggered fermions being in the wrong universality class. 
\end{abstract}
\maketitle
\section{ Introduction and main argument}
\label{sec:intro}
The authors of \cite{Fodor:2016zil}, see also \cite{Fodor:2017gtj, Fodor:2017nlp}, report on the absence of an infrared (IR) fixed point for QCD with twelve flavors, thus concluding that $N_f\!=\!12$ lies below the lower edge, $N_f^c$, of the QCD conformal window. The authors of \cite{Hasenfratz:2016dou, Cheng:2014jba} reach instead the opposite conclusion for the same theory with the same method. In \cite{Hasenfratz:2017qyr, Hasenfratz:2017mdh} this discrepancy has been attributed to staggered fermions being in the wrong universality class. 
The conclusion in \cite{Fodor:2016zil} and \cite{Hasenfratz:2016dou, Cheng:2014jba} is based on a lattice study of the finite volume gradient flow of the expectation value of the Lagrangian density, denoted as  $\langle E(t)\rangle$, and used to define the renormalized gauge coupling as proposed in \cite{Fodor:2012td}. Specifically, 
the composite operator $E(t,x)$ is the Lagrangian density, ${\mbox{Tr}}G^2$, evolved to ``flow time'' $t\!\geqslant\!0$ in Euclidean metric: 
\ba
\label{eq:E}
E(t,x) &=& \frac{1}{2}{\mbox{Tr}}G_{\mu\nu}G_{\mu\nu}\nonumber\\
G_{\mu\nu}(t,x)&=&\partial_\mu B_\nu - \partial_\nu B_\mu +ig [B_\mu,B_\nu]\, ,
\ea
where $B_\mu(t,x)$ is the $t$-evolved gauge field solution of the gradient-flow equation:
\be
\label{eq:GF}
\dot{B}_\mu=D_\nu G_{\nu\mu}\,,
\ee
with initial condition $B_\mu(t,x)|_{t=0}\!=\!A_\mu(x)$, and $A_\mu(x)$ the bare gauge field. The dot in Eq.~(\ref{eq:GF}) stands for the derivative with respect to $t$ and $D_\mu\!=\!\partial_\mu\!+i\!g[B_\mu, \cdot\,]$. 
The flow time, $\sqrt{t}$, acts as the smearing radius for the gauge field $B_\mu(t,x)$.
Throughout this work we write quantities in the canonical normalization of perturbation theory, i.e., after the gauge field has been rescaled by the gauge coupling.    

In  \cite{Luscher:2010iy} it was shown that at one loop in perturbation theory at infinite volume one has:
\be
\label{eq:Luscher1loop}
t^2\langle E(t)\rangle\sim \,1+\beta_0g^2(\mu)\log{(8t\mu^2)} \,= \frac{g_{1l}^2(t)}{g^2(\mu)},
\ee
where $\beta_0$ is the universal, i.e., renormalization-scheme independent one-loop coefficient of the beta function $\beta(g)\!=\!-\!\beta_0g^3\!-\!\beta_1g^5+O(g^7)$, $g_{1l}(t)$ is the one-loop running coupling at the scale $\sqrt{t}$ and, consistently with the canonical normalization in Eq.~(\ref{eq:E}), we divided out the constant factor $g^2(\mu)$ from Eq.~(2.32) in \cite{Luscher:2010iy}.

Equation~(\ref{eq:Luscher1loop}) implies that, after the standard one-loop multiplicative renormalization of ${\mbox{Tr}}G^2$, $t^2\langle E(t)\rangle$ is one-loop finite.
The finiteness of all correlation functions of multiplicatively renormalized operators at positive flow time $t$ has then been proved to all orders in perturbation theory in \cite{Luscher:2011bx}. In other words, the gradient flow does not modify the renormalization properties of the theory. 

Subsequently, the authors of \cite{Fodor:2012td} have  suggested to define a would-be nonperturbative running coupling, ${g_{wb}}(t)$, by means of the renormalized operator $t^2\langle E(t)\rangle$. Consistently with Eq.~(\ref{eq:Luscher1loop}), we write the definition in \cite{Fodor:2012td} as follows:
\be
\label{eq:flowVinfty}
g^2(\mu ) t^2\langle E(t)\rangle = {g}_{wb}^2(t)\frac{3(N^2-1)}{128\pi^2}\,,
\ee
and we discuss its finite volume version in Sec.~\ref{sec:FV}. 
\subsection{Main argument}
\label{sec:mainarg}
Our main observation is that Eq.~(\ref{eq:flowVinfty}) defines 
a would-be running coupling:
\be
\label{eq:gwb}
 g_{wb}^2(t)\propto g^2(\mu)t^2\langle E(t)\rangle =g^2(\mu){\cal G}(g(t))Z(\sqrt{t}\mu,g(\mu))
\ee
 that not only depends on a renormalization-group invariant (RGI) function, ${\cal G}$, of a true RGI running coupling, $g(t)$, but also depends on the multiplicative 
 renormalization factor, $Z$, of $t^2\langle E(t)\rangle$, according to the solution in Eq.~(\ref{eq:Esolution}) of the Callan-Symanzik equation derived in Sec.~\ref{sec:RGE}. 

A true running coupling, $g(t)$ in Eq.~(\ref{eq:gwb}), is defined by the renormalization of QCD and, therefore, is RGI differently from $g_{wb}(t)$, though, obviously, 
renormalization-scheme dependent. 

According to Eq.~(\ref{eq:gwb}), the would-be beta function $\beta_{wb}({g}_{wb}^2(t))\!=\!-d{g}_{wb}^2(t)/d\log\sqrt{t}$ reads:
\ba
\label{eq:wouldbe-beta}
&&\hspace{-0.3cm}\beta_{wb}({g}_{wb}^2(t))= {g}_{wb}^2(t)\left (\beta(g(t))\frac{d\log{\cal G}(g(t))}{dg(t)}+\gamma(g(t))\right )\nonumber\\
&&\hspace{-0.3cm}= {g}_{wb}^2(t)\left \{\frac{\beta(g(t))}{g(t)}   \left ( g(t)\frac{d\log{\cal G}(g(t))}{dg(t)} - 1  \right )\! +\!\beta'(g(t))\right \}
\ea
where $\beta(g(t))\!=\!-dg(t)/d\log\sqrt{t}$ is a true beta function for a true running coupling $g(t)$. A true beta function $\beta(g(t))\!=\!-\!\beta_0g(t)^3\!-\!\beta_1g(t)^5+O(g(t)^7)$ is defined by the renormalization of QCD and, therefore, it is a RGI function of $g(t)$ only, with renormalization-scheme independent coefficients $\beta_0$ and $\beta_1$. We refer to the latter property as the two-loop universality of a true beta function. As a consequence of the RG invariance of a true beta function, in addition to the universal zero of asymptotic freedom at vanishing coupling,  $\beta(g(t))$ in the conformal window has a renormalization-scheme dependent zero whose existence is universal. 
In Eq.~(\ref{eq:wouldbe-beta}) $\gamma(g(t))\!=\!-d\log Z(\sqrt{t}\mu,g(\mu))/d\log\sqrt{t}$ is the anomalous dimension of 
$t^2\langle E(t)\rangle$
and, importantly:
\be
\label{eq:gamma_exact}
\gamma(g)=g\frac{\partial}{\partial g}\left (\frac{\beta (g )}{g}   \right )
=\beta '(g )-\frac{\beta (g )}{g}\,,
\ee
with $\beta'$ the derivative with respect to $g$, see e.g. \cite{Bochicchio:2013tfa,NunesdaSilva:2016jfy}.

The would-be beta function, $\beta_{wb}$ in Eq.~(\ref{eq:wouldbe-beta}), depends on the anomalous dimension, $\gamma(g)$, in a way that spoils its proportionality to a true beta function. 

Most relevant is what happens at a true nontrivial IR fixed point, which occurs at a necessarily nonvanishing coupling, $g_*$, defined by a zero of a true beta function, $\beta(g_*)\!=\!0$,
with anomalous dimension at the IR fixed point given by:
\be
\label{eq:gammaIR}
\gamma_*=\beta'(g_*)\geqslant 0\,,
\ee
according to Eq.~(\ref{eq:gamma_exact}).
As a true coupling, $g(t)$, approaches $g_*$, the Callan-Symanzik Eq.~(\ref{eq:RGEmu_E}) in Sec.~\ref{sec:RGE} implies the asymptotic conformal scaling: 
\be
\label{eq:IRscaling}
g_{wb}^2(t)\propto g^2(\mu)t^2\langle E(t)\rangle \sim t^{-\gamma_*/2}~~~\mbox{as}~~t\rightarrow\infty\, .
\ee
Thus, $g_{wb}^2(t)$ vanishes asymptotically in the IR for $\gamma_*\!>\!0$.
Intuitively,  Eq.~(\ref{eq:IRscaling}) can be understood by dimensional reasoning due to the fact that $t$ is the only scale in the massless infinite volume theory. 
Correspondingly, Eq.~(\ref{eq:wouldbe-beta}) implies the IR asymptotic scaling for $\beta_{wb}$:
\be
\label{eq:IRscaling_betawb}
\beta_{wb}({g}_{wb}^2(t))\sim\gamma_*g_{wb}^2(t)\sim\gamma_*t^{-\gamma_*/2}
~~~\mbox{as}~~t\rightarrow\infty\,,
\ee
which again vanishes asymptotically in the IR for $\gamma_*\!>\!0$.  

Moreover, $\beta_{wb}$  is renormalization-scheme dependent beyond one loop -- since so is  $\gamma(g)$ -- and thus violates the two-loop universality of a true beta function. As a consequence, $\beta_{wb}$ may develop spurious zeroes due to 
renormalization-scheme dependent cancellations that may occur between the two terms in Eq.~(\ref{eq:wouldbe-beta}) at finite $t$, both in the confined phase and the conformal window. In fact, a spurious zero must exist in presence of an IR fixed point if $\gamma_*\!>\!0$, see Sec.~\ref{sec:comp}.

These features have crucial consequences, further discussed in Sec.~\ref{sec:comp}, if our aim is to discriminate between the confined phase,  
$N_f\!<\!N_f^c$, and the conformal window in the interval $N_f^c\!\leqslant\! N_f\!<\! N_f^{AF}$, where a nontrivial IR fixed point of a true running coupling occurs.  

Only in a neighborhood of the ultraviolet (UV) asymptotically free fixed point the would-be running coupling, $g_{wb}^2(t)$ in Eq.~(\ref{eq:flowVinfty}), is asymptotic to a true running coupling, $g^2(t)$.  
In fact, as shown in Sec.~\ref{sec:RGE}, the ultraviolet universal asymptotic behavior of  $g_{wb}^2(t)$   reads:
\ba
\label{eq:Eoneloop}
&&\hspace{-0.25truecm}
g_{wb}^2(t)\propto g^2(\mu)t^2\langle E(t)\rangle\sim g^2(\mu) \left (  \frac{g(t)}{g(\mu)}\right )^{\frac{\gamma_0}{\beta_0}} 
= g^2(\mu)
\nonumber\\
&&
\hspace{-0.25truecm}
 \times \left (  \frac{g(t)}{g(\mu)}\right )^2\sim 
\frac{1}{\beta_0\log\frac{1}{t\Lambda_{QCD}^2 }  } \left (1-  \frac{\beta_1\log\log\frac{1}{t\Lambda_{QCD}^2 }}{\beta_0^2\log\frac{1}{t\Lambda_{QCD}^2 }
   }
\right )\,\,\,
\ea
where, importantly, $\gamma_0\!=\! 2\beta_0$ \cite{Bochicchio:2013tfa,Bochicchio:2013eda} is the universal one-loop coefficient of  $\gamma(g)\!=\!-\gamma_0g^2\! -\!\gamma_1g^4\!+\!O(g^6)$, while $\gamma_1$ is renormalization-scheme dependent, see e.g. \cite{Bochicchio:2013tfa}, 
and we used the universal asymptotic expression for the running coupling  \cite{Bochicchio:2013tfa,Bochicchio:2013eda} for $\sqrt{t}\!\ll\!\Lambda_{QCD}^{-1}$, with $\Lambda_{QCD}$ the RGI  scale of the asymptotically free theory that exists both in the confined phase, $N_f\!<\!N_f^c$, and in the conformal window, even if the latter is deconfined. Equation~(\ref{eq:Luscher1loop})  
is the perturbative version of  Eq.~(\ref{eq:Eoneloop}). 
\subsubsection{ Incompatibility of  the ansatz in \cite{Fodor:2012td} with the QCD fundamental properties}
\label{sec:comp}
In contrast with Eq.~(\ref{eq:IRscaling}), the authors of \cite{Fodor:2012td} postulate that, for $N_f\!<\!N_f^{AF}$ in the conformal window,  the would-be running coupling, ${g}_{wb}(t)$ in Eq.~(\ref{eq:flowVinfty}), attains its would-be IR fixed point value, ${g}_{wb}^*$, defined by a zero of the associated would-be beta function:
\be
\label{eq:IRscaling-nogamma-nozero}
\beta_{wb}({g}_{wb}^{*2})=0\,.
\ee
Thus ${g}_{wb}^*$ should necessarily be nonvanishing at a nontrivial IR fixed point, so that:
\be
\label{eq:IRscaling-nogamma}
g_{wb}^2(t)\propto g^2(\mu)t^2\langle E(t)\rangle \sim {g}_{wb}^{*2}\neq 0~~~\mbox{as}~~t\rightarrow\infty\,.
\ee
Clearly, Eqs.~(\ref{eq:IRscaling-nogamma-nozero}) and   (\ref{eq:IRscaling-nogamma}) are incompatible with Eqs.~(\ref{eq:IRscaling}) and (\ref{eq:IRscaling_betawb}) for  nonvanishing $\gamma_*$. 
They agree only for $\gamma_*\!=\!0$, which according to Eq.~(\ref{eq:gammaIR})
can only occur at a true IR fixed point, $g_*$, if a true beta function has a multiple zero. 
This happens at the upper edge, $N_f^{AF}$, where the theory is IR free, hence $g_*\!=\!0$, but  there is no reason for it to happen at a nontrivial IR fixed point. 
In fact, the vanishing of $\gamma_*$ is inconsistent with the QCD prediction 
-- exact for large $N$ and $N_f$ in perturbation theory -- of a nonvanishing and positive
$\gamma_*\!=\!16\epsilon^2/225(1+O(\epsilon))$ \cite{Banks:1981nn} at 
a nontrivial IR fixed point in a neighborhood of $N_f^{AF}\!=\!(11/2)N$, with $\epsilon\!=\!11/2\!-\!N_f/N\!\ll\!1$.
The vanishing of $\gamma_*$ is also inconsistent with the absence, proved in \cite{NunesdaSilva:2016jfy}, of an UV-IR fixed-point merging at the lower edge of the SQCD conformal window, which would imply a double zero of a true beta function.  Moreover, a nonvanishing $\gamma_*$ is consistent with the features \cite{NunesdaSilva:2016jfy} of the exact beta function of large-$N$ QCD in the Veneziano limit  
\cite{Bochicchio:2013aha,Bochicchio:2008vt,Bochicchio:2012bj}. In the same large-$N$ framework, both in QCD and SQCD, 
$\gamma_*$ is strictly positive for $N_f\!<\!N_f^{AF}$ and increases  along the IR fixed point curve as $N_f$ decreases towards $N_f^c$ \cite{NunesdaSilva:2016jfy}.
Thus, $t^2\langle E(t)\rangle$ in Eq.~(\ref{eq:IRscaling}) vanishes asymptotically with a power-law rate
that increases as $\gamma_*$ increases from $N_f^{AF}$ down to $N_f^c$.

A fact relevant for lattice studies is that the IR asymptotics with $\gamma_*\!>\!0$ in Eq.~(\ref{eq:IRscaling})  and the  UV asymptotics  in Eq.~(\ref{eq:Eoneloop}) 
imply that $t^2\langle E(t)\rangle$ must attain a maximum
at some finite $t$. This maximum is a spurious zero of $\beta_{wb}$ in Eq.~(\ref{eq:wouldbe-beta}) that hence must occur for each $N_f$ with $\gamma_*\!>\!0$ in the conformal window. However, as already mentioned, additional spurious zeroes of $\beta_{wb}$ may occur both in the confined phase and the conformal window, thus implying that the observation of a maximum of  $t^2\langle E(t)\rangle$, i.e., a spurious zero of $\beta_{wb}$, is necessary but not sufficient to establish that a theory is in the conformal window. 
If $\gamma_*\!>\!0$, the asymptotic conformal scaling in Eq.~(\ref{eq:IRscaling}), or equivalently Eq.~(\ref{eq:IRscaling_betawb}), is the unique physical property whose observation is necessary and sufficient to establish that a theory is in the conformal window by means of the gradient-flow method.

This concludes our main argument. 
In Sec.~\ref{sec:RGE} we derive Eqs.~(\ref{eq:gwb}), (\ref{eq:IRscaling}) and (\ref{eq:Eoneloop}).
We generalize the results to the finite volume case in Sec.~\ref{sec:FV}. In Sec.~\ref{sec:results} we discuss current lattice results, in particular \cite{Fodor:2016zil,Fodor:2017gtj, Fodor:2017nlp,Hasenfratz:2016dou, Cheng:2014jba}. Section \ref{sec:zeromodes} is a note on gauge zero modes. Sec.~\ref{sec:staggered} comments on the universality class of staggered fermions.  We conclude and suggest new strategies in Sec.~\ref{sec:mainconc}.  
\subsection{RGE solution and UV/IR asymptotics}
\label{sec:RGE}
In order to understand the physics in a neighborhood of $g\!=\!0$ in the UV and $g\!=\!g_*$ in the IR, it is convenient to use the general solution of the Callan-Symanzik equation in the Euclidean coordinate representation for the two-point correlator of the 
composite operator ${\mbox{Tr}}G^2$ in QCD derived in \cite{Bochicchio:2013tfa,Bochicchio:2013eda}, with  $N_f\!<\!N_f^{AF}$  massless flavors in the fundamental representation.
An analogous equation with an analogous solution will then imply the asymptotic scaling with the flow time $t$ of the smeared one-point function given by $\langle E(t)\rangle$. 

The renormalized two-point correlator $G^{(2)}(x)\!\equiv\! \langle{\mbox{Tr}}G^2(x){\mbox{Tr}}G^2(0)\rangle$ at nonzero separation $x\!\neq\!0$ -- that avoids contact terms and guarantees multiplicative renormalization -- obeys the Callan-Symanzik equation \cite{Bochicchio:2013tfa,Bochicchio:2013eda}:
\be
\label{eq:RGEmu}
\left (x\cdot\frac{\partial}{\partial x} +\beta(g)\frac{\partial}{\partial g} +2D +2\gamma(g)\right ) G^{(2)}\left (x, {\mu}, g(\mu)\right )=0,
\ee
where $D\!=\!4$ is the canonical dimension of ${\mbox{Tr}}G^2$ in four spacetime dimensions and $\gamma(g)\!=\!-\partial\log Z/\partial\log\mu$ is the anomalous dimension of  ${\mbox{Tr}}G^2$. 
 The general solution of Eq.~(\ref{eq:RGEmu}) thus reads \cite{Bochicchio:2013tfa,Bochicchio:2013eda}:
\ba
\label{eq:RGEsol}
\hspace{-0.5cm}G^{(2)}\left (x, {\mu}, g(\mu)\right )
&=&  \frac{1}{x^{2D}} {\cal G}\left ( g(x)\right ) Z^2\left (x\mu ,g(\mu)\right )
\nonumber\\
&=&   \frac{1}{x^{2D}} {\cal G}\left ( g(x)\right ) 
e^{2\int_{g(\mu)}^{g(x)}~\frac{\gamma(g)}{\beta(g)}dg       }\,,
\ea
where $Z$ is the renormalization factor of ${\mbox{Tr}}G^2$,
${\cal G}\left ( g(x)\right ) $ is a RGI function of $g(x)$, and ${\cal G}\left ( g(x)\right )\!\sim\! 1$ \cite{Bochicchio:2013tfa,Bochicchio:2013eda} as $g(x)\!\to\! 0$, for $G^{(2)}(x)$ does not vanish at lowest order in perturbation theory.

The ultraviolet universal asymptotic behavior of Eq.~(\ref{eq:RGEsol}) 
is then obtained using ${\cal G}\left ( g(x)\right )\!\sim\! 1$, the one-loop anomalous dimension, $\gamma(g)\!=\!-\gamma_0g^2\!+\!O(g^4)$, and the two-loop beta function, $\beta(g)\!=\! -\beta_0g^3\!-\!\beta_1g^5\!+\!O(g^7)$, leading to \cite{Bochicchio:2013tfa,Bochicchio:2013eda}:
\ba
\label{eq:RGEsol_pert}
&&\hspace{-0.2cm}G^{(2)} (x)
\sim
\frac{1}{x^8}
\left (\frac{g^2(x)}{g^2(\mu)}\right )^{\frac{\gamma_0}{\beta_0}}
= \frac{1}{x^8}\left (\frac{g^2(x)}{g^2(\mu)}\right )^2
 \nonumber\\
&&\hspace{-0.2cm}\sim\frac{1}{x^8} 
 \left ( 
\frac{1}{\beta_0\log\frac{1}{x^2\Lambda_{QCD}^2 }  }
 \left (1-  \frac{\beta_1\log\log\frac{1}{x^2\Lambda_{QCD}^2 }}{\beta_0^2\log\frac{1}{x^2\Lambda_{QCD}^2 }
   }
\right )
\right )^2 
\ea
where we used the universal asymptotic running coupling
\cite{Bochicchio:2013tfa,Bochicchio:2013eda} for $x\!\ll\!\Lambda_{QCD}^{-1}$ and, 
 importantly, $\gamma_0\!=\! 2\beta_0$ \cite{Bochicchio:2013tfa,Bochicchio:2013eda}.

For a given $N_f$ in the conformal window, the asymptotic infrared behavior of $G^{(2)}(x)$ in a neighborhood of $g\!=\!g_*$ as $x\!\rightarrow\!\infty$ is obtained by expanding Eq.~(\ref{eq:RGEsol}) around $g\!=\!g_*$, with $\beta(g)\!=\!\gamma_*(g\!-\!g_*)+O((g\!-\!g_*)^2)$ and $\gamma_*\!=\!\beta'(g_*)$ according to Eq.~(\ref{eq:gammaIR}):
 \be
\label{eq:RGEsol_IR}
G^{(2)}(x)
\sim 
\frac{1}{x^{8+2\gamma_*}}\,.
\ee
 This would also describe the scaling of $G^{(2)}(x)$ for an exactly conformal theory with anomalous dimension $\gamma_*$. 

We now note that, since the gradient-flow equation does not change the renormalization properties of the theory, and we assume multiplicative renormalizability according to \cite{Luscher:2011bx} for $t\!\neq\!0$, we can write a Callan-Symanzik equation analogous to Eq.~(\ref{eq:RGEmu}) for the smeared one-point function 
 $G^{(1)}(t)\!\equiv\!\langle E(t)\rangle$:
\be
\label{eq:RGEmu_E}
\left (\sqrt{t}\frac{\partial}{\partial\sqrt{t}} +\beta(g)\frac{\partial}{\partial g} +4+\gamma(g)\right ) G^{(1)}\left (\sqrt{t}, {\mu}, g(\mu)\right )=0\, ,
\ee
whose general solution reads:
\be
\label{eq:Esolution}
G^{(1)}\left (\sqrt{t}, {\mu}, g(\mu)\right )
= \frac{1}{{t^2}} {\cal G}\left ( g(t)\right ) Z(\sqrt{t}\mu,g(\mu))\,,
\ee
where
${\cal G}\left ( g(t)\right ) $ is a RGI function of $g(t)$ and $Z$ is defined in Eq.~(\ref{eq:RGEsol}), by replacing $x$ with $\sqrt{t}$. 
The UV asymptotic solution is then Eq.~(\ref{eq:Eoneloop}) and the IR asymptotic solution is Eq.~(\ref{eq:IRscaling}), in full analogy with  Eq.~(\ref{eq:RGEsol_pert}) and  Eq.~(\ref{eq:RGEsol_IR}), respectively.
\subsection{Finite volume}
\label{sec:FV}
Eq.~(\ref{eq:flowVinfty}) has been generalized to the finite volume case, specifically a Euclidean spacetime box of size $L^4$ in  \cite{Fodor:2012td}:
\be
\label{eq:flow}
 g^2(\mu)t^2\langle E(t)\rangle = {g}_{wb}^2(t)\frac{3(N^2-1)}{128\pi^2}(1+\delta(c))\,,
\ee
which again defines a would-be running coupling, ${g}_{wb}(t)$, that now 
 runs with $\sqrt{8t}\!=\!cL$,  with the dimensionless ratio $c\!=\!\sqrt{8t}/L$ held fixed according to \cite{Fodor:2012td}. 
The term $\delta(c)$ in Eq.~(\ref{eq:flow}) contains the finite volume corrections in the UV to the infinite volume case in Eq.~(\ref{eq:flowVinfty}). Different choices of $c$ correspond to different renormalization schemes according to \cite{Fodor:2012td}. 
In Eq.~(\ref{eq:flow}) the UV limit is taken as $t\!\rightarrow\! 0$
with $L$ held fixed, thus $c\!\rightarrow\! 0$ and $\delta(c)\!\rightarrow\! 0$. The IR limit needs further discussion because it is taken in lattice simulations as $L\!\rightarrow\!\infty$ with $c\!>\!0$ held fixed. 

The definition of the would-be finite-volume running coupling in Eq.~(\ref{eq:flow}) is not RGI as the infinite volume one in Eq.~(\ref{eq:flowVinfty}). 
Moreover, it involves additional difficulties that we summarize as follows:
\begin{itemize}
\item[1)]  When the IR scale $L$ is introduced and  $t^2\langle E(t)\rangle$ is measured for some $g(t)\!\neq\!g_*$ and a fixed ratio $c$ of $\sqrt{t}$ (the ``UV scale'') and $L$ (the ``IR scale''), finite volume corrections modify the Callan-Symanzik Eq.~(\ref{eq:RGEmu_E}) in a way that is not presently under theoretical control.  Indeed,  the term $\delta(c)$ only guarantees the correct UV asymptotic behavior in Eq.~(\ref{eq:flow}). 
\item[2)] A smaller $c$, i.e., a smaller ratio of the UV/IR scales, implies smaller finite volume corrections, allowing for a better determination in the conformal window of the maximum of $t^2\langle E(t)\rangle$ that precedes the asymptotic scaling in Eq.~(\ref{eq:IRscaling}).
 However, Eq.~(\ref{eq:IRscaling}) is only retrieved as $t\!\to\!\infty$ in the infinite volume theory, i.e., only after the limit $L\!\to\!\infty$ has been taken with $t$ held fixed, thus $c\!\to\!0$. Current lattice studies have instead performed the limit $L\!\to\!\infty$ with $c\!>\!0$ held fixed. As a consequence,   
 the corresponding IR limit $t\!\to\!\infty$ still contains finite volume corrections.  
\item[3)] If $\gamma_*\!>\!0$ in the conformal window, 
a maximum of $t^2\langle E(t)\rangle$ must occur at infinite volume and finite $t$, thus for $c\!=\!0$, and its location and height also depend on $N_f$. However, for $c\!>\!0$, even the existence of this maximum is unclear because of the combined effect of renormalization-scheme dependence and finite volume corrections that never vanish according to 2).  
\end{itemize}
These features render the determination of the lower edge of the conformal window with the gradient-flow method rather difficult. 
\section{Lattice results}
\label{sec:results}
In order to determine whether the $N_f\!=\!12$ theory is in the conformal window, present lattice studies and in particular \cite{Fodor:2016zil,Fodor:2017gtj, Fodor:2017nlp,Hasenfratz:2016dou, Cheng:2014jba} have looked for a zero of the discrete and finite volume version of the would-be beta function $\beta_{wb}$ in Eq.~(\ref{eq:wouldbe-beta}), specifically, $\sigma (s,L)\!=\!({g}_{wb}^2(sL)\!-\!{g}_{wb}^2(L))/\log{(s^2)}$, with $s\!>\!1$ the size of the discrete step and ${g}_{wb}(sL)$ defined by Eq.~(\ref{eq:flow}), with $c\!>\!0$ held fixed. 

The difficulty is that, in presence of a true IR fixed point with $\gamma_*\!>\!0$,  the approach of $\beta_{wb}$ to the asymptotic scaling in 
Eq.~(\ref{eq:IRscaling_betawb})  is affected by nonuniversal finite volume corrections, see 1) in Sec.~\ref{sec:FV}. Moreover, $\beta_{wb}$ may develop spurious zeroes whose existence is affected by the choice of renormalization scheme and finite volume corrections, see 3) in Sec.~\ref{sec:FV}.

As explained in 2) of Sec.~\ref{sec:FV}, despite the limit $L\!\to\!\infty$ of $\sigma (s,L)$ is taken in \cite{Fodor:2016zil,Fodor:2017gtj, Fodor:2017nlp,Hasenfratz:2016dou, Cheng:2014jba}, the finite volume corrections in $\sigma (s,L)$ are not removed since $c\!>\!0$ is held fixed. Therefore, $\sigma (s,L)$ does not recover the infinite volume $\beta_{wb}$.

A spurious zero of $\sigma (s,L)$ may have presumably been observed in \cite{Hasenfratz:2016dou, Cheng:2014jba} for the $N_f\!=\!12$ theory, being it the maximum of $t^2\langle E(t)\rangle$ at infinite volume and finite $t$ in the conformal window, or another spurious zero at finite volume in the confined or the conformal phase, see Sec.~\ref{sec:FV}. Then, 
the numerical differences for $\sigma(s,L)$ between \cite{Fodor:2016zil,Fodor:2017gtj, Fodor:2017nlp} and \cite{Hasenfratz:2016dou, Cheng:2014jba}  may be attributed to different choices of the renormalization scheme due to different choices of $c$, and different  finite volume corrections, in a region of the lattice parameter space that is not where the conformal scaling in Eq.~(\ref{eq:IRscaling}) and Eq.~(\ref{eq:IRscaling_betawb}) sets in, if the theory is in the conformal window. 

Thus, the analysis in \cite{Fodor:2016zil,Fodor:2017gtj, Fodor:2017nlp,Hasenfratz:2016dou, Cheng:2014jba} cannot yet establish conclusively the nature of the $N_f\!=\!12$ theory, see Sec.~\ref{sec:mainconc} for a suggestion on how to improve this analysis. 
\subsection{Note on gauge zero modes}
\label{sec:zeromodes}
 In the studies \cite{Fodor:2016zil,Fodor:2017gtj, Fodor:2017nlp, Hasenfratz:2016dou, Cheng:2014jba} Eq.~(\ref{eq:flow}) contains via $\delta(c)$ the contribution of zero momentum gauge modes.
It is desirable to remove the gauge zero modes by employing twisted or mixed Dirichlet-Neumann boundary conditions because they enhance finite volume effects and possibly cutoff effects in two ways. 
 First, already in the UV regime the gauge zero modes contribute with dominant $1/L^4$ contributions to $\delta (c)$ \cite{Fodor:2012td} as compared to the exponentially suppressed contributions of the nonzero modes. Even more
importantly, they induce a modified perturbative expansion  of $\langle E(t)\rangle$ that now contains odd powers of the coupling according to \cite{Fodor:2012td}. These terms are genuine finite volume effects that vanish as $c\!\to\!0$, but they now appear at $O(g^3)$ instead of $O(g^4)$, thus enhanced, and even in the UV they violate the universal asymptotics of the next-to-leading logarithms 
 in Eq.~(\ref{eq:Eoneloop}). As a consequence, they also augment the violation of two-loop universality with respect to the infinite volume  $\beta_{wb}$ in Eq.~(\ref{eq:wouldbe-beta}).   
\subsection{Staggered fermions}
\label{sec:staggered}
In an attempt to explain the disagreement between \cite{Fodor:2016zil,Fodor:2017gtj, Fodor:2017nlp} and \cite{Hasenfratz:2016dou,Cheng:2014jba}, it was claimed in \cite{Hasenfratz:2017qyr, Hasenfratz:2017mdh} that staggered fermions are in the wrong universality class \cite{Hasenfratz:2017qyr, Hasenfratz:2017mdh} and, therefore, cannot be used to study the IR fixed point in the conformal window \cite{Hasenfratz:2017qyr, Hasenfratz:2017mdh}, whereas they could well be used to study the UV asymptotics of QCD \cite{Hasenfratz:2017qyr, Hasenfratz:2017mdh}. 
We observe that any deviation of the staggered fermion formulation from the continuum theory is due to discretization effects which are  genuine UV effects. 
Thus, any statement on the universality class of staggered fermions that has been made in the context of asymptotically free and confining QCD holds true in the infrared of the conformal window. 

The numerical differences observed in \cite{Hasenfratz:2017qyr,Hasenfratz:2017mdh} for $N_f\!=\!10, 12$  between $\sigma(s,L)$ with domain-wall fermions and $\sigma(s,L)$ with staggered fermions may be explained by our analysis in Sec.~\ref{sec:results}  and may have no physical implication, nor they may indicate the failure of a specific lattice fermion formulation. 

We suggest that an additional 
 source of numerical difference may be that staggered fermions are massless in \cite{Fodor:2016zil,Fodor:2017gtj, Fodor:2017nlp,Hasenfratz:2016dou,Cheng:2014jba} and \cite{Hasenfratz:2017qyr,Hasenfratz:2017mdh}, whereas domain-wall fermions have $m_{res}\!\neq\! 0$ in \cite{Hasenfratz:2017qyr,Hasenfratz:2017mdh}. Indeed, in the latter case, the observed strong-coupling bulk transition or crossover -- possibly breaking the exact chiral symmetry of the weakly coupled  theory in the conformal window -- is likely to affect all measured quantities in its vicinity. Instead, a chiral symmetry breaking bulk transition cannot occur in the massless staggered fermions case, because the massless limit taken at finite volume prevents from recovering spontaneous chiral symmetry breaking at infinite volume, due to the well known noncommutativity of the limits  $m\!\to\! 0$ and $V\!\to\!\infty$ \cite{Leutwyler:1992yt,Banks:1979yr}. 
\vspace{0.1cm}
\section{Conclusions}
\label{sec:mainconc}
We have shown that the would-be running coupling defined by means of the gradient-flow method applied to ${\mbox{Tr}}G^2$   is not RGI, but carries an anomalous dimension. Therefore, 
future gradient flow studies of $t^2\langle E(t)\rangle$ for large-$N_f$ QCD-like theories should first demonstrate to be able to access the asymptotic scaling region 
of $t^2\langle E(t)\rangle$ in Eq.~(\ref{eq:IRscaling}) due to its anomalous dimension, as opposed to the would-be scaling in Eq.~(\ref{eq:IRscaling-nogamma}) proposed in \cite{Fodor:2012td}.
Specifically, one should reproduce the QCD prediction  of the upper edge at $N_f^{AF}$ \cite{Banks:1981nn,NunesdaSilva:2016jfy}, working in a neighborhood $N_f\!\lesssim\!N_f^{AF}$, where a nonvanishing $\gamma_*$ distinguishes crucially Eq.~(\ref{eq:IRscaling}) from Eq.~(\ref{eq:IRscaling-nogamma}). After this preliminary test of the method -- essential to show that one is able to discriminate between the conformal and the confined phase -- one may then proceed towards lower $N_f$, until $N_f^c$ is crossed.  

Yet, the determination of  $N_f^c$ with this method remains difficult, 
due to a choice of renormalization scheme with fixed $c\!>\!0$ that does not allow to get rid of finite volume corrections, see  Sec.~\ref{sec:FV}. The elimination of gauge zero modes may help reducing the latters, see Sec.~\ref{sec:zeromodes}. 

Finally, we observe that 
a RGI operator that directly probes the existence of an IR fixed point is the trace of the energy-momentum tensor in QCD with massless quarks,
$T_\mu^\mu\!=\!(\beta(g)/g){\mbox{Tr}}G^2$, which vanishes at the fixed point because the beta function vanishes. 
The gradient flow $t^2\langle T_\mu^\mu(t)\rangle\!=\!{\cal G}_T(g(t))$ is a function of $g(t)$ only, whose beta function, $-d{\cal G}_T/d\log\sqrt{t}\!=\! \beta(g)d{\cal G}_T/dg$, also vanishes at a true IR fixed point because $\beta(g)$ vanishes. 
 Another quantity of interest may be the two-point correlator of the RGI operator $g^2{\mbox{Tr}}G\tilde{G}$, with $\tilde{G}$ the dual of $G$, see \cite{Bochicchio:2013tfa,Bochicchio:2013eda} for a thorough analysis of its renormalization properties.
The gradient flow $t^2\langle T_\mu^\mu(t)\rangle$, or the $n$-point correlation functions 
of $T_\mu^\mu$ or $g^2{\mbox{Tr}}G\tilde{G}$ may offer an alternative to $t^2\langle E(t)\rangle$ for lattice studies of the conformal window, together with a mandatory study of the phase transitions of the lattice system. 
\section*{Acknowledgments}
We thank M. Bochicchio for inspiring discussions and for reading the manuscript. 
\balance
\bibliography{references.bib}
\end{document}